\documentclass[superscriptaddress,reprint,onecolumn]{revtex4-1}

\usepackage{amsmath}
\usepackage{amsfonts}
\usepackage{amssymb}
\usepackage{amsthm}
\usepackage{amsxtra}
\usepackage{soul} 
\usepackage{fancyhdr}
\usepackage{mathrsfs}
\usepackage{graphicx}
\usepackage{dcolumn}
\usepackage{bm}
\usepackage[colorlinks = true, citecolor = blue, linkcolor = blue]{hyperref}

\newcommand{\be}{\begin{equation}}
\newcommand{\ee}{\end{equation}}
\newcommand{\ben}{\begin{eqnarray}}
\newcommand{\een}{\end{eqnarray}}

\newcommand{\cor}[1]{\left[ #1 \right]}
\newcommand{\pare}[1]{\left( #1 \right)}
\newcommand{\key}[1]{\left\{ #1 \right\}}
\newcommand{\ket}[1]{|#1\rangle}
\newcommand{\bra}[1]{\langle #1|}

\newcommand{\op}[2]{\ket{#1}\!\bra{#2}}

\newcommand{\mean}[1]{\langle #1 \rangle}

\DeclareMathOperator{\Tr}{Tr}

\begin{document}
\title{An atomic symmetry-controlled thermal switch}

\author{Daniel Manzano}
\email{manzano@onsager.ugr.es}
\affiliation{Engineering Product Development, Singapore University of Technology and Design, 8 Somapah Road, 487372 Singapore, Singapore}
\affiliation{Department of Chemistry, Massachusetts Institute of Technology, 77 Mass Ave.,  02139 Cambridge, USA}
\affiliation{Department of Electromagnetism and Matter Physics. University of Granada, Av. Fuentenueva, 18007 Granada, Spain }
\author{Elica Kyoseva}
\email{elica\_kyoseva@sutd.edu.sg}
\affiliation{Engineering Product Development, Singapore University of Technology and Design, 8 Somapah Road, 487372 Singapore, Singapore}

\date{\today}


\begin{abstract}
We propose a simple diatomic system trapped inside an optical cavity to control the energy flow between two thermal baths. Through the action of the baths the system is driven to a non-equilibrium steady state. Using the Large Deviation theory we show that the number of photons flowing between the two baths is dramatically different depending on the symmetry of the atomic states. Here we present a deterministic scheme to prepare symmetric and antisymmetric atomic states with the use of external driving fields, thus implementing an atomic control switch for the energy flow.
\end{abstract}

\maketitle
\section*{Introduction} 

Understanding how energy flows and how to control this flow has recently attracted considerable attention \cite{dubi:rmp11}. It further prompted the study of transport phenomena in quantum systems including the analysis of fundamental laws like Fourier's Law \cite{znidaric:jsm10,sun:epl10,manzano:pre12,asadian:pre13} and the second law of thermodynamics \cite{brandao:pnas}.  Furthermore, it was shown that quantum effects can be used to enhance important thermodynamical processes. Examples include quantum thermal machines \cite{linden:prl10} and solar cells \cite{scully:prl10,creatore:prl13,wang:njp14}. Another exciting field of research is thermal phenomena and design of electronic and thermal devices at the nanoscale, where several experiments demonstrated electronic switches at the atomic scale \cite{fuechsle:nn12,park:nature02,schirm:nn13}. Notably, a two-terminal switch to control the charge transport in a junction based on the reversible rearrangement of single atoms was shown in Ref. \cite{schirm:nn13}. 

One of the most promising physical platforms to study quantum transport phenomena is coupled optical cavities with one or more atoms trapped per site. In order to design different topologies, optical cavities can be linked via fibers \cite{cirac:prl97,vanenk:prl99,kyoseva:njp12} with an effective interaction given by Bose-Hubbard Hamiltonians \cite{jaksch:prl98}. These cavities are suitable to model transport in systems with next-neighbor interactions with lineal or network topologies \cite{kyoseva:njp12}. The control capability  of optical cavities make them a very useful tool to study transport phenomena. Applications include the study of quantum transport in different dimensions \cite{luo,manzano:arxiv}, and the simulation of noise-assisted transport \cite{caruso:pra11}. 

Recently, Bu\v{ca} and Prosen showed that certain symmetries in an open quantum system lead to different nonequilibrium steady states \cite{buca:njp12} (see also \cite{popkov:njp13,baumgartner:jpa08}). These steady states are classified according to the symmetry operator spectrum. The multiplicity of steady states leads to  different expected currents. This effect can be used to design a symmetry-controlled quantum switch to govern the energy current passing through the system by simply selecting its initial state \cite{manzano:prb14,thigna:prb14}. Subsequently, the role of symmetry in energy transfer was studied in both transient \cite{plenio2a,plenio2b} and steady state scenarios \cite{manzano:po13}. However, harnessing quantum symmetry to control the energy flux in a steady state scenario was considered only in toy models \cite{manzano:prb14}.

In this paper we propose a feasible design of an {\it atomic symmetry-controlled thermal switch}. Our design comprises a chain of cavities, coupled, at the two ends, to two different temperature baths, $T_1$ and $T_2$, while the middle cavity is doped with two laser-driven atoms. The thermal baths drive the system out of equilibrium and it evolves to a steady state scenario, where there is a finite energy current flowing from the hot to the cold bath. We show how a pair of atoms trapped in the middle cavity can be used to control the energy current through the system up to four orders of magnitude. The control is implemented, in a deterministic fashion, by switching between the  symmetric and antisymmetric atomic state manifolds with the use of a laser driving field. 

\section*{Results and Discussion}
 Our system comprises a chain of three optical cavities coupled to thermal baths at temperatures $T_1$ and $T_2$, as it is shown in Figure (1). The middle cavity is doped with two three-state atoms addressed by external laser fields, while the cavity photons hop between neighboring sites at a rate $J$. The evolution of the system is given by a Markovian master equation \cite{breuer_02} in the form ($\hbar=1$),
\be
\dot{\rho}=- i \pare{H\rho - \rho H } + \sum_{i = 1}^4 \mathcal{L}^{\text{at}}_i \rho +  \sum_{c = \{l, r\}} \mathcal{L}^{\textnormal{th}}_c \rho , 
\label{eq:me}
\ee
where $H = H^{\text{ctrl}} + H^{\text{hop}}$ is the rotating-wave approximation (RWA) Hamiltonian accounting for the control atom-laser interaction and the hopping of photons, $H^{\text{hop}} =   J\left(a_l^\dagger a_2 +  a_2^\dagger a_r  + H.c. \right)$. Here $a_{l/2/r}$ the anhilitation operator of photons of the left/middle/right cavity, while $\mathcal{L} \rho = L \rho L^{\dagger} -\frac{1}{2} \key{L^\dagger L,\rho}$ are Lindblad superoperators
, which describe the interaction of the system with the incoherent channels. There are four decay channels associated with the atomic states [see Figure (1) right], which are accounted for by $\mathcal{L}_i^{\text{at}}$ with $i=\{1,\dots,4 \}$, and two decay channels $\mathcal{L}_c^{\text{th}}$, with $c = \{l, r\}$, associated with the thermal bath coupled to the left and right cavity, respectively. The RWA is taken for simplicity, but it does neither affect the multiplicity of steady states nor the control capacity of the system.

\begin{figure}
\includegraphics[scale=0.5]{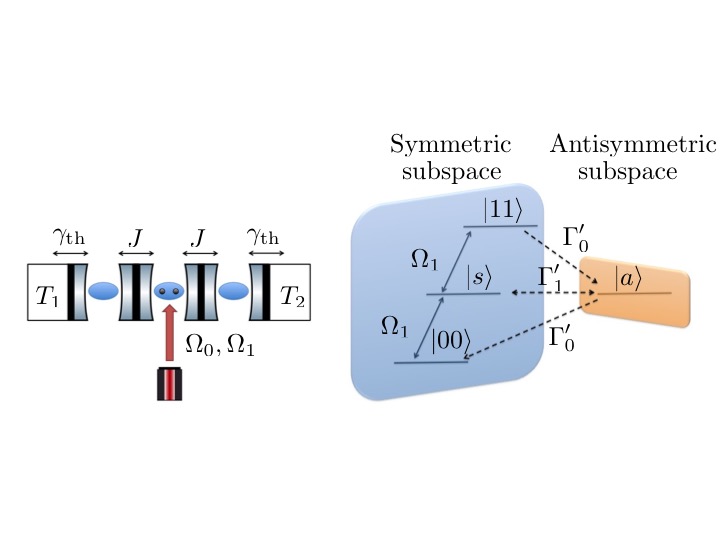}
\caption{Left: A schematic of a symmetry-controlled switch to control the current between two thermal baths, $T_1$ and $T_2$. Right: Energy diagram of the symmetric and antisymmetric atomic subspaces, which modify the energy current between the baths up to four orders of magnitude, and the possible transitions between them.}
\end{figure}

To realize a controlled deterministic switching between atomic states with different symmetry we adopt a scheme, where the quantum state preparation is heralded by macroscopic quantum jumps \cite{metz:prl06,metz:pra07}. We assume that two identical three-state $\Lambda$ atoms are coupled to the middle cavity's mode $a_2$ on the $|0\rangle \leftrightarrow |e\rangle$ transition with a coupling strength $g$. Furthermore, the $|1\rangle \leftrightarrow |e\rangle$ and $|0\rangle \leftrightarrow |1\rangle$ transitions are driven by two external laser fields with Rabi frequencies $\Omega_0$ and $\Omega_1$, respectively. If the excited states $|e\rangle$ are far off-resonant, i.e. $\Omega_1 <g,\Gamma,\Omega_0 \ll \Delta$, where $\Delta$ and $\Gamma$ are the detuning and decay rate of the excited states, then they can be adiabatically eliminated. Thus, the state space of the diatomic system is reduced to contain only four states, three of which $|00\rangle$, $|11\rangle$, and $\ket{s}=(\ket{01}+\ket{10})/\sqrt{2}$, are symmetric, while $\ket{a}=(\ket{01}-\ket{10})/\sqrt{2}$ is antisymmetric, as displayed in Figure (1) right. The system is then equivalent to a four level system. Then, the control Hamiltonian is given by (see Methods section)
\be
H^{\text{ctrl}} = \tfrac{1}{\sqrt{2}} \Omega_1 \left( \op{00}{s} + \op{s}{11} + H.c. \right)  +  g'  \pare{  \op{00}{s}a_2^{\dagger}  +     \op{s}{11}a_2^{\dagger}  + H.c  }    
+ \Delta' \left( \op{00}{00} - \op{11}{11} \right),
\label{eq:hamiltonian_total}
\ee
with $\Delta' = -\tfrac{g^2}{\Delta} a_2^{\dag}a_2 - \tfrac{\Omega_0^2}{4\Delta}$, and $g' = -\tfrac{\Omega_0 g}{\sqrt{2}\Delta}$. 

The transitions between the symmetric and antisymmetric atomic states occur as a result of two decay channels,
\be
L^\text{at}_{1} = \sqrt{ \Gamma'_0} \left[ \op{00}{a} -\op{a}{11}  \right],\quad 
L^\text{at}_{2} = \sqrt{\tfrac{\Gamma'_1}{2}} \left[  \op{s}{a} + \op{a}{s}  \right],
\label{eq:reset_atoms}
\ee
while there are two more decay channels which link the symmetric states, $L^\text{at}_{3}=\sqrt{\Gamma'_0} \left[ \op{00}{s} +\op{s}{11}  \right] $ and $L^\text{at}_{4}=\sqrt{\tfrac{\Gamma'_1}{2}} \left[ \op{a}{a} +\op{s}{s} +2\op{11}{11}   \right]$, see Figure (1) right. Here, the decay rates are modified by the external driving fields and are explicitly given by $\Gamma'_{0(1)} = \tfrac{\Gamma_{0(1)}\Omega_0^2}{4 \Delta^2}$, with $\Gamma_0 + \Gamma_1 = \Gamma$, \cite{metz:prl06,metz:pra07}. Note that, there is mixing between the symmetric and antisymmetric atomic subspaces only when $\Omega_0 \neq 0$.

\begin{figure}
\includegraphics[scale=0.4]{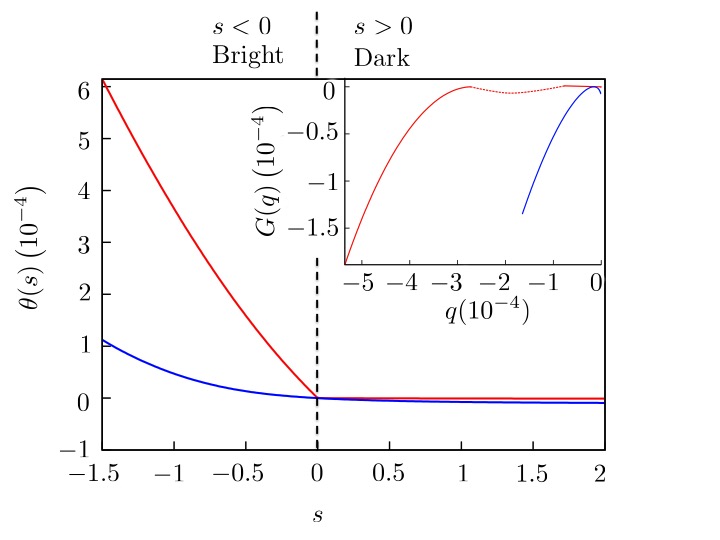}
\caption{Large deviation functions $\theta(s)$ (main) and $G(q)$ (inset). In both plots the blue line represents the state with the laser on  ($\Omega_1=0.005g,\; \Omega_0=g$) and the red line with the laser off  ($\Omega_1=\Omega_0=0$). The remaining parameters are   
$\Delta=75g,\;\Gamma_0=\Gamma_1,\;\Gamma=g,\;J=10^{-3}g,\; n_1=0.005,\;n_2=10^{-6}$. The dashed line in $G(q)$ sketches the non-convex regime that cannot be calculated from $\theta(s)$.}
\end{figure}

The presence of the baths leads to the creation and destruction of cavity photons in the left and right cavities, which we model by the following incoherent channels,  
\ben
L^\text{th}_{l_1} &=&  \sqrt{\gamma_\text{th} n_1 } a_l, \,\,\,  L^\text{th}_{l_2} =  \sqrt{\gamma_\text{th}( n_1 +1)}\, a_l^\dagger, \nonumber \\ 
L^\text{th}_{r_1} &=& \sqrt{\gamma_\text{th} n_2 } a_r, \,\,\, L^\text{th}_{r_2}=  \sqrt{\gamma_\text{th}( n_2 +1)} \, a_r^\dagger,
\label{eq:lindblad_thermal}
\een
where subscripts 1 and 2 correspond to creation and destruction, respectively, $\gamma_\text{th}$ is the interaction rate, $n_{1(2)}=1/[\exp(\omega/(k_B T_{1(2)} )) - 1]$ is the temperature-dependent mean excitation photon number at the resonance frequency in the respective bosonic bath, and $k_{\text{B}}$ is the Boltzmann's constant~\cite{breuer_02}.

To analise the transport properties of our system we use a Large Deviation approach \cite{pre:esposito06a,pre:esposito06b,garrahan:prl10,manzano:prb14} . The key elements of this technique are the large deviation functions $\theta(s)$ and $G(q)$ that account for the statistic of photons interchanged between the system and one of the baths.  These functions determine the values for long times of both the moment-generating function, $Z_s(t) \sim \exp \cor{t \theta(s}$, and the probability of having a certain value $q$ of the current, $P_q(t)\sim \exp \cor{t \theta (s)}$ (see Methods). The moments of the current distribution can be calculated directly from the large deviation function $\theta(s)$ by the expression $\mean{q^\alpha}=\left. - \frac{\partial^\alpha \theta(s) }{\partial s^\alpha} \right| _{s=0}$.
 
In Figure 2 the moment-generator LDF $\theta(s)$ is displayed for the cases when the lasers are driving the atoms ($\Omega_{0,1}\ne 0$) and when they are off ($\Omega_{0,1}=0$). When the lasers are off there are two different steady states in the system. This is reflected in the kink of $\theta(s)$ at $s=0$. The non-analyticity of $\theta(s)$ can be interpreted as a dynamical phase transition, and it is a consequence of the existence of more than one steady state with different activity. If the atoms are in one of the symmetric states $\{|00\rangle, |11\rangle, |s\rangle \}$, then they facilitate the transfer of photons to the cold reservoir $T_1$. This is described by $\theta(s)$ in the region with $s<0$. However, if the atoms are in the antisymmetric state $|a\rangle$, which is a dark state for the system, then they cannot interact with the cavity photons, thereby reducing the energy transfer through the cavities. This scenario is depicted in the LDF $\theta(s)$ for the regime $s>0$. 

The multiplicity of the steady states is a consequence of the symmetries of the system \cite{manzano:prb14}. If $\Omega_0=0$ (laser off) the unitary operator $\pi$ that interchanges the states of the two atoms ($\pi\ket{00}=\ket{00},\; \pi\ket{s}=\ket{s},\; \pi\ket{11}=\ket{11}, \;\pi\ket{a}=-\ket{a}$) commutes with both the Hamiltonian and all of the Lindblad operators, $\cor{\pi,H}=\cor{\pi,L^{\text{at}}_i}=\cor{\pi,L^{\text{th}}_c}=0,\; \forall (i,c).$ This leads to a degeneracy of the Liouvillian operator and to multiple steady states, which can be labeled by the eigenvalues of the operator $\pi$ \cite{buca:njp12}. When the laser is on, meaning that $\Omega_0\ne 0$, the degeneracy is broken, because the operator $L_{1}^{\text{at}}$ from Eq. \eqref{eq:reset_atoms} does not commute with $\pi$ because $\cor{\pi, L^{\text{at}}_i}=2L^{\text{at}}_i $ is equal to zero if and only if $L^{\text{at}}_i=0$, that corresponds with $\Omega_0= 0$. In this case there is only one steady state that mixes the symmetric and antisymmetric manifolds of the atomic states. Because of this, there is no dynamical phase transition and $\theta(s)$ and its derivative are analytical for all values of $s$, including zero, as it is displayed in Figure 2.

We recover the large deviation function $G(q)$ from $\theta(s)$ by a Legendre transformation. The non-analyticity of the derivative of $\theta(s)$ leads to a non-convex regime in $G(q)$ that cannot be directly inferred by the Legendre transformation, as it is sketched in the inset of Figure 2. The probability of having $Q$ events in a long time $t$ can be calculated as $P_Q(t)= \exp [tG(Q/t)]$.

To compare the efficiency of the symmetric and antisymmetric steady states we calculate the mean number of photons transferred to the reservoir per unit time $\mean{q}$ directly from $\theta(s)$ according to $\mean{q}=\left. -\frac{\partial \theta(s)}{\partial s} \right|_{s=0}.$ If there are more than one steady states, the mean flux for the more and less active steady states are given by $\mean{q}_\text{max}=\left. -\frac{\partial \theta(s)}{\partial s} \right|_{\substack{s\to0\\ s<0}}$ and $\mean{q}_\text{min}=\left. -\frac{\partial \theta(s)}{\partial s} \right|_{\substack{s\to0\\ s>0}}$, respectively.  We calculate the ratio between the maximum and minimum flux $\alpha=\mean{q}_\text{max}/\mean{q}_\text{min}$. In Figure 3 we plot $\alpha$ as a function of the hopping parameter $J$ and we find that $\alpha$ is strongly dependent on the value of $J$, up to three orders of magnitude. The control capacity of the system increases dramatically when $J$ decreases, and vice versa. The reason is that $J$ modifies the coupling of the atoms to the cavity modes -- the smaller $J$ is the stronger the atoms are coupled to the photons--  which leads to an increase in the control efficiency of the current.

\begin{figure}
\includegraphics[scale=0.2]{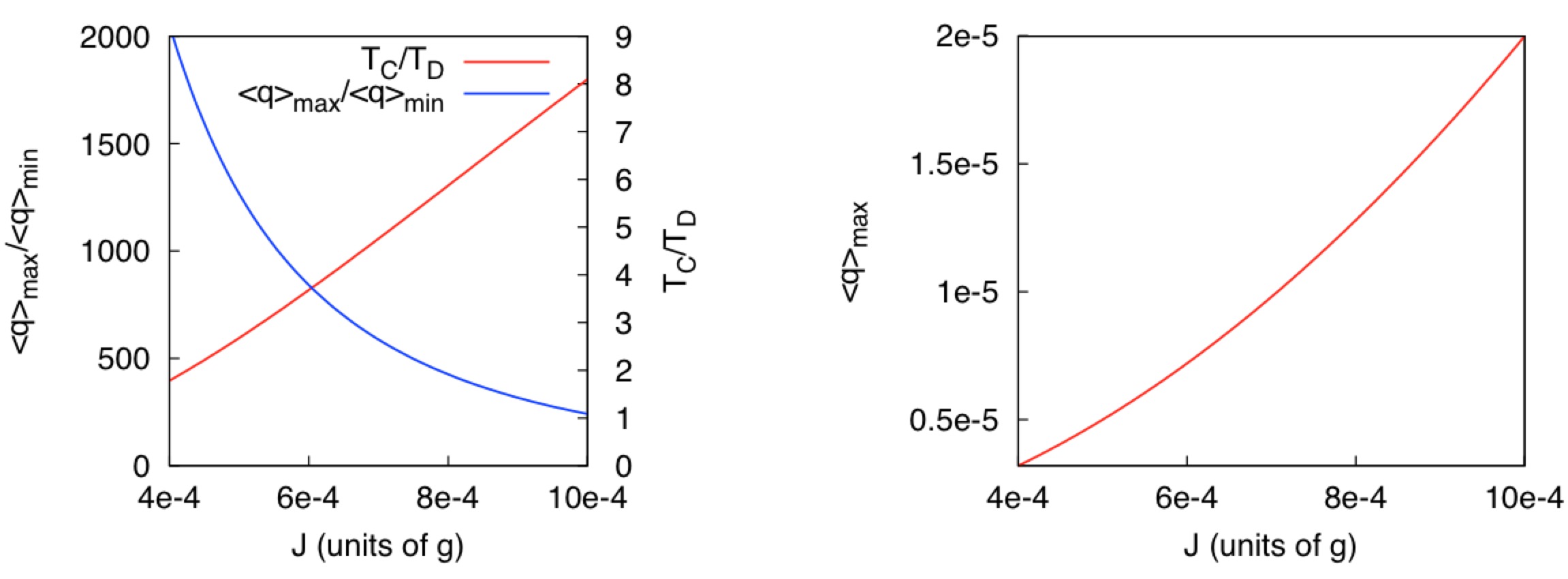}
\caption{Left, blue: Ratio between maximum and minimum flux as a function of the hopping parameter $J$. We have assumed no driving of the atoms $(\Omega_1=\Omega_0=0)$. Left, red: Ratio between the mean length of a dark period and the mean time between the absorption of photons in a bright period ($\Omega_1=0.0025g,\; \Omega_0=g$). Right: Total flux for the bright state ($\Omega_1=0.0025g,\; \Omega_0=g$) as a function of the hopping parameter $J$. The flux is expressed in atomic units and it corresponds to $\sim 10^{13}$ photons  per second. In all plots the remaining parameters are $\Gamma_0=\Gamma_1,\;\Gamma=g,\; n_1=0.1,\;n_2=0$. }
\end{figure}

To control the switch we need to be able to distinguish between the symmetric and antisymmetric (dark) states. For our parameter selection the dark state current is close to zero, and the probability of photon absorption for this state is negligible. This means that if one photon is absorbed by the cold reservoir we are in a bright state with a very high probability. To be able to detect it we need to ensure that the time between absorption of photons in the bright state is smaller than the mean length of the dark period. The probability of transition from a dark into a light period is $\Gamma'_0 + \frac{1}{2} \Gamma'_1$, and the mean length of a dark period equals \cite{metz:pra07},
\be
T_D=\frac{1}{2\Gamma_0 + \Gamma_1} \frac{8 \Delta^2}{\Omega^2_0}.
\ee
The mean time between photon absorption can be calculated as the inverse of the mean flux $T_C=1/\mean{q}$. As the mean flux is an average of the bright and dark fluxes we can consider it as a lower bound of the flux when the atoms are in the bright manifold. In Figure 3 (left) we can see that the ratio $T_C/T_D$ increases with the hopping between cavities $J$. There is a competition between the capacity to distinguish the bright and dark periods that require a high current, and the ratio between the bright and dark flux that is higher for smaller fluxes. Nonetheless, for a broad range of the parameters the time between photon absorption in the bright state is long enough, compared to the mean length of the dark period and the photon flux of the bright state is more than two orders of magnitude larger than the photon flux in the dark state. 

The maximum flux, $\mean{q}_{\text{max}}$, is displayed in Figure 3 (right) as a function of the cavities hopping parameter $J$. Due to the small value of the coupling parameter $J$,  the calculated values for $\mean{q}_{\text{max}}$ are also small but, notably, they are measurable. To assess this, we performed a simple calculation and obtained a $\mean{q}_{\text{max}}\sim  10^{-5}$ (in atomic units) that corresponds to a flux of the order of $10^{13}$ photons per seconds. Given such light intensity for the bright period, it will be easily distinguished from the dark period when the flux is several orders of magnitude smaller.  The proposed design has several potential applications. It can be used as an energy current switch, an atomic memory and also an entangling scheme due to the indirect measurement of the atomic state through the measurement of the current. This is particularly useful for memory and entanglement generation, as the antisymmetric state of the atoms is a maximally entangled state. Thus, by monitoring the energy current we can control the separability properties of the atoms. The readout of the atomic state can be performed without directly interacting with the atoms themselves, as their state affects the macroscopic light flux. The proposed system can also be extended to realize better control by placing more pairs of atoms in the cavities. By connecting these cavities with optical fibers we can create networks of cavities \cite{kyoseva:njp12}, each of them acting as a switch, thereby increasing the flux control.

In this paper we have presented a realistic model of an atomic thermal switch based on coupled optical cavities. We showed that by placing two laser-driven atoms in an optical cavity it is possible to control the energy current trough the system when it is driven out of equilibrium by the action of two thermal baths at different temperatures. 

The control capacity of the system depends on the system parameters, increasing when the hopping between cavities decreases. For small values of the hopping parameter between the cavities the current of the bright state can be three orders of magnitude bigger than the current of the dark state. This can be used also as a quantum memory as the state of the atoms that can be measured without distorting the atoms, just by measuring the current  between the full system and the thermal bath.  
The importance of this design relies on the simplicity of manipulating an internal degree of freedom of a system in comparison with modifying the system parameters. Similar symmetry-controlled switches can be designed for many different quantum setups such as quantum dots, optical lattices and trapped ions. 

\section*{Methods}
{\bf Calculation of the Master Equation.-}
The symmetry control of our system is given by a pair of $\Lambda$ atoms at the second cavity driven by a laser. This scheme was first proposed to create entangled states by macroscopic quantum jumps \cite{metz:prl06,metz:pra07}. The Hamiltonian of the system can be decomposed in the way $H=H^{\text{ctrl}} + H^{\text{hop}}$, were $H^{\text{ctrl}}$ is the part of the Hamiltonian that models the interaction between the atoms and the laser field and $H^{\text{hop}}$ is the photon hopping between cavities. The atoms are trapped in the cavity in a way such that the effect of the laser is the same in both of them. The coupling strength of the $0-2$ transition to the cavity field is $g$, and the laser Rabi frequencies for the $\ket{0}-\ket{1}$ and $\ket{1}-\ket{2}$ transitions are $\Omega_0$ and $\Omega_1$,

\be
H^{\text{ctrl}} = \sum_{i=1}^2 \pare{\frac{\Omega_1}{2} \ket{1}_i \bra{2}_{i} + \frac{\Omega_0}{2} \ket{0}_i \bra{1}_i + H.c.  }
+ \sum_{i=1}^2 \pare{ g \ket{0}_i \bra{2}_i  a_2^{\dagger} +H.c. } + \Delta \ket{2}_i\bra{2}_i 
\ee
where $a_2^\dagger$ is the bosonic ladder operator that creates and destroys a photon at cavity $2$. It is assumed that $\Omega_0 <g,\Gamma,\Omega_1 \ll \Delta$.  As the detunning $\Delta$ is bigger than the other parameters of the system the excited atomic states can be adiabatically eliminated. The system is then equivalent to a four level system. As both atoms are equally affected by the laser, it is convenient to use the Bell basis for describing the state of the two atoms, with $\ket{s}=(\ket{01}+\ket{10})/\sqrt{2}$ and $\ket{a}=(\ket{01}-\ket{10})/\sqrt{2}$. The control Hamiltonian becomes 

\be
H^{\text{ctrl}} = \tfrac{1}{\sqrt{2}} \Omega_1 \left( \op{00}{s} + \op{s}{11} + H.c. \right)  +  g'  \pare{  \op{00}{s}a_2^{\dagger}  +     \op{s}{11}a_2^{\dagger}  + H.c  }    
+ \Delta' \left( \op{00}{00} - \op{11}{11} \right),
\label{eq:hamiltonian_total}
\ee
with $\Delta' = -\tfrac{g^2}{\Delta} a_2^{\dag}a_2 - \tfrac{\Omega_0^2}{4\Delta}$, and $g' = -\tfrac{\Omega_0 g}{\sqrt{2}\Delta}$. 
The Lindblad operators that correspond to the atoms incoherent channels are 

\ben
L^\text{at}_{1} &=& \sqrt{ \Gamma'_0} \left[ \op{00}{a} -\op{a}{11}  \right],\quad 
L^\text{at}_{2} = \sqrt{\tfrac{\Gamma'_1}{2}} \left[  \op{s}{a} + \op{a}{s}  \right], \nonumber\\
L^\text{at}_{3} &=& \sqrt{\Gamma'_0} \left[ \op{00}{s} +\op{s}{11}  \right]  ,\quad 
L^\text{at}_{4} = \sqrt{\tfrac{\Gamma'_1}{2}} \left[ \op{a}{a} +\op{s}{s} +2\op{11}{11}   \right]
\een
where  $\Gamma'_{0(1)} = \tfrac{\Gamma_{0(1)}\Omega_0^2}{4 \Delta^2}$.

{\bf Large deviation approach.-} To analyse the statistics of the current flowing between the system and the thermal reservoirs we use a full-counting statistics method~\cite{pre:esposito06a,pre:esposito06b,garrahan:prl10,manzano:prb14}. For simplicity, we focus only on the current between the system and the $T_2$ reservoir. We first introduce the reduced density matrix $\rho_Q(t)$, which is the projection of the full density matrix onto the subspace with $Q$ photons interchanged between the system and the $T_2$ bath. It is not normalised and the probability of having a certain value of the current $Q$ in a period of time $t$ is given by $P_Q(t)=\Tr[\rho_Q(t)]$. This probability scales for long times following a large deviation principle, $P_Q(t)\sim \exp [tG(q)]\; (q=Q/t)$, where $G(q)$ is the LDF of the current. 
 
The direct calculation of $G(q)$ is complicated, but it becomes simpler after a change of ensemble. We make a Laplace transform on $\rho_Q(t)$ and obtain a new reduced density matrix $\rho_s=\sum_Q \rho_Q(t) \exp (-s Q) $, where we have introduced a {\it counting field} $s$, that is the conjugate field of the number of photons $Q$. The fields $Q$ and $s$ are dynamical variables with the same relation as the thermodynamical variables pressure and volume. This procedure is formally equivalent to the thermodynamic transformation between the canonical and microcanonical potentials. The utility of this transformation comes from the fact that the evolution of the reduced density matrix $\rho_s$ unravels into a set of equations in the form
\be
 \dot{\rho}_s 
= - i \pare{H \rho_s  - \rho_s  H} + \sum_{i=1}^4 \mathcal{L}^{\text{at}}_i \rho_s   + \mathcal{L}^\text{th}_{l_1} \rho_s  +\mathcal{L}^\text{th}_{l_2} \rho_s + 
\widetilde{\mathcal{L}}^\text{th}_{r_1} \rho_s  +  \widetilde{\mathcal{L}}^\text{th}_{r_2} \rho_s
\equiv \mathcal{W}_s \rho_s,
\ee
where $\widetilde{\mathcal{L}}^\text{th}_{r_1} \rho=e^{-s} L^{\dagger}_{r_1} \rho L^{\phantom{\dagger}}_{r_1}-\frac{1}{2} \key{L^{\phantom{\dagger}}_{r_1}L^{\dagger}_{r_1},\rho}$ and $\widetilde{\mathcal{L}}^\text{th}_{r_2} \rho=e^s L^{\dagger}_{r_2} \rho L^{\phantom{\dagger}}_{r_2}-\frac{1}{2} \key{L^{\phantom{\dagger}}_{r_2}L^{\dagger}_{r_2},\rho}$, and they account for the introduction of the counting field $s$ in the right current.

The Laplace transform fulfills a large deviation principle in the form $Z_s(t) \sim \exp [t \theta(s)]$, with  $\theta(s)$ being a large deviation function related to $G(q)$ by a Legendre-type transform $\theta(s)=\max_q[G(q)-s q]$. Here we have introduced $Z_s(t)$, which is the generating function of the current momenta. The LDF $\theta(s)$ corresponds to the eigenvalue of the superoperator $\mathcal{W}_s$ with the largest real part. The existence of more than one steady states with different currents leads to a non-analytic behavior of the LDF $\theta(s)$ at $s=0$ \cite{manzano:prb14}. 

To calculate the statistical properties of the current we need to diagonalize the superoperator $\mathcal{W}_s$. However, this is a numerically intractable problem as the dimension of our system is infinite due to the infinite number of photon modes. To overcome this challenge, we adopt a low photon number approximation, where the number of photons in the system is small, which can be realized if the temperatures of both baths are very small and the frequency of the laser $\Omega_0$ is also small. In this regime we assume that there can be at most one photon in the total system. Under this assumption, the Hilbert space required to describe the system is equivalent to the Hilbert space of a four qubit-system. Note that, we make the low photon number approximation only for numerical tractability. Our scheme will operate as a current switch independent of the number of photons in the system.

\section*{Acknowledgements}
This work has been supported by an SUTD-MIT International Design Centre (IDC) Grant, Project No. IDG 31300102. D.M. acknowledges  financial support from MIT-SUTD program, and the Junta de Andaluc'a and EU Project TAHUB/II-148 (Program ANDALUCêA TALENT HUB 291780). E.K. acknowledges  financial support by an SUTD Start-Up Research Grant, Project No. SRG EPD 2012029 and by the project COPQE by H2020 Marie Scklodowska-Curie Actions Individual Fellowships with Project No. 705256.

\section*{Author Contributions Statement}
E.K. designed the system and performed the analytical analysis. D.M. carried out the numerical simulations and data analysis. E.K. and D.M. contributed equally to the manuscript preparation.
 
\section*{Competing Financial Interests statement}
The authors declare no competing financial interests.

\end{document}